\documentclass{appolb}

\usepackage{graphicx}

\newcommand{\tabref}[1]{Table~\ref{#1}}

\renewcommand{\Re}{\mathop{\mathrm{Re}}}
\renewcommand{\Im}{\mathop{\mathrm{Im}}}

\begin{document}
\title{Charmonium excitation functions in $\bar p A$ collisions%
\thanks{Presented at Excited QCD 2018, Kapaonik, Serbia}%
}
\author{Gy. Wolf, G. Balassa, P. Kov\'acs, M. Z\'et\'enyi,
\address{\vspace*{-0.4cm} Wigner RCP, Budapest, 1525 POB 49, Hungary}
\\[0.6cm]
Su Houng Lee
\address{\vspace*{-0.4cm} Department of Physics and Institute of Physics and Applied Physics, Yonsei University, Seoul 120-749, Korea}
}
\maketitle
\begin{abstract}
  We study the excitation function of the low-lying charmonium state:
  $\Psi$(3686) in $\bar p$ Au collisions taking into account their
  in-medium propagation. The time evolution of the spectral functions
  of the charmonium state is studied with a BUU type transport model.
  We calculated the excitation function of $\Psi$(3686) production and show
  that it is strongly effected by the medium. The energy regime will be
  available for the PANDA experiment. 
\end{abstract}
  
\section{Introduction}

One of the most important quantities in the phenomenology of strong
interaction is the gluon condensate. We know its value in vacuum, actually its
numerical value was obtained by the sum rule for $J/\Psi$ \cite{CharmSumRule}.
However, in medium we only know its first derivative with respect to the
density obtained from the trace anomaly relation. Its value in matter,
therefore, would be of great importance.
While the masses of hadrons consisting of light quarks changes mainly because
of the (partial) restoration of chiral symmetry those made
of heavy quarks are sensitive mainly to the changes of the non-perturbative
gluon dynamics manifested through the changes in the  gluon
condensates \cite{Luke:1992tm,Klingl:1998sr}. In the low density
approximation the gluon condensate is expected to be reduced by $5-7
\%$ at normal nuclear density \cite{Meissner:1995,SHLee_Charm}. Therefore,
the masses of the
charmonium states -- which can be considered as color dipoles in color
electric field -- are shifted downwards because of the second order
Stark effect \cite{SHLee_Charm,LeeKo,Friman:2002fs}. Moreover, since
the $D$ meson loops contribute to the charmonium self\,-\,energies and
they are slightly modified in\,-\,medium, these modifications generate
further minor contributions to the charmonium in\,-\,medium mass
shifts \cite{LeeKo}.

In antiproton induced reactions a large number of charmed states are expected to
be produced, creating a laboratory for studying the gluon condensate.
The observation of charmonium in vacuum is an important goal of
the PANDA collaboration at the future FAIR complex. Besides that, the
observation of charmonium production in matter is also a part of the PANDA
program.
The $\bar p A$ reactions are best suited to observe charmed particles
in nuclear matter, since in this case the medium is much simpler than the one
created in a heavy ion collisions. (The other advantageous channel is the pion
induced reaction with energies of 12-15 GeV pion energy, such beam is available
at JPARC, Tokai, Japan.) Furthermore, the two main background
contribution to the dilepton yield in the charmonium region, namely
the Drell-Yan and the ``open charm decay'' are expected to be
small. There are only a few energetic hadron\,-\,hadron collisions that
can produce heavy dileptons via the Drell-Yan process.  In the open
charm decay the $D$ mesons decay weakly. Consequently, the $e^+$ and $e^-$
are usually accompanied by
the $K$ and $\bar K$ mesons. Therefore, not very far above the
threshold, the production of electron\,-\,positron pairs via the open
charm decay with large invariant mass is energetically suppressed. 

The spectral functions of the J/$\Psi$, $\Psi$(3686), and $\Psi$(3770) vector
mesons are expected to be modified in a strongly interacting environment.
In our transport model of the BUU type the time evolution of 
single-particle distribution functions of various hadrons are evaluated within
the framework of a kinetic theory. We studied in our previous work what is
the effect of the
in-medium modification of the charmonium spectral function on the dilepton
yield well above threshold \cite{WolfCharm_PhysLett}. We have learned that the
most promissing candidate to observe the in-medium modifications is the
$\Psi$(3686) meson. In this paper we study the effect of the medium on the
$\Psi$(3686) meson around threshold, the excitation function with vacuum and
with in-medium spectral functions.

\section{Off-shell transport of broad resonances}

Our transport model, originally valid in the few GeV energy range, was
recently upgraded to higher energies of up to 10 AGeV \cite{WolfCharm_PhysLett}.
We use a statistical model \cite{Balassa:2017pgf} to calculate the unknown cross sections, such as $\bar p p \to J/\psi \pi$, or $\bar p p \to D\bar D$.
We apply energy independent charmonium
absorption cross sections for every hadron $4.18$~mb for $J/\Psi$
and $7.6$~mb for $\Psi(3686)$ and $\Psi(3770)$ according to Ref.
\cite{Linnyk:2006ti}. In $\bar p A$ collisions at relativistic energies,
charmonium absorption does not play such an important role as at
ultrarelativistic energies, since the hadron density is much less here.
It should be noted that the decay of the charmonium states is
handled perturbatively.

If we create a particle in a medium with in-medium mass, it should regain its
vacuum mass during the propagation, by the time it reaches the vacuum. 
We can describe the in-medium properties of particles with
an ``off-shell transport''. These equations are
derived by starting from the Kadanoff\,-\,Baym equations
\cite{Baym:1961zz} for the Green's functions of the
particles. Applying first\,-\,order gradient expansion after a Wigner
transformation \cite{Cassing-Juchem00,Leupold00}, one arrives at a
transport equation for the retarded Green's function. To solve
numerically the  off-shell transport equations one may exploit the
test\,-\,particle ansatz for the retarded Green's function
\cite{Cassing-Juchem00,Leupold00}.

The equations of motion of the test-particles 
have to be supplemented by a collision term
which couples the equations of the different particle species.
It can be shown \cite{Leupold00} that 
the collision term has the same form as in the
standard BUU treatment.

In our calculations we  employ a simple 
form of the self-energy of a vector meson $V$:
\begin{eqnarray} \label{areal}
{\rm \Re} \Sigma^{ret}_V & = & 2 m_V \Delta m_V \frac{n}{n_0},\\
\label{aimag}
{\rm \Im} \Sigma^{ret}_V & = & -m_V (\Gamma^{vac}_V + \Gamma_{coll}).
\end{eqnarray}
Eq.~(\ref{areal}) describes a ''mass shift'' 
$\Delta m = \sqrt{m_V^2+\Re\Sigma_V^{ret}}-m_V\approx  \Delta m_V \frac{n}{n_0}$.
The imaginary part incorporates a vacuum
width $\Gamma^{vac}_V$ term and a collisional broadening term having the form
\begin{equation}
  \Gamma_{coll} = \frac{v \sigma \rho}{\sqrt{(1-v^2)}},
\end{equation}
where $v=|\vec{p}|/m$ is the velocity of the particle in the local rest
frame, $\sigma$ is the total cross section of the particle colliding
with nucleons and $\rho$ is the local density. The parameters $\Delta
m_V$ are taken from \cite{LeeKo} and are given in \tabref{Tab:param_dmV}.

\begin{table}[th]
  \caption{\label{Tab:param_dmV} Charmonium mass shift parameter
    values taken from \cite{LeeKo}. In $\Delta m_V$ the first term
    result from the second order Stark\,-\,effect, while the second
    term emanates from the D\,-\,meson loops.}
  \begin{tabular}{cc}
    Charmonium type ($V$) &  $\Delta m_V$\\
    \hline
    $J/\Psi$     & $-8+3$~MeV  \\
    $\Psi(3686)$ & $-100-30$~MeV \\
    $\Psi(3770)$ & $-140+15$~MeV \\
  \end{tabular}
\end{table}
The first values in \tabref{Tab:param_dmV}
come from the second order Stark\,-\,effect (which depends on the
gluon condensate), while the second ones emanate from the D\,-\,meson
loops.

If a meson is generated at normal density its mass is distributed in
accordance with the in-medium spectral function. If the meson
propagates then its mass will be modified according to $\Re\Sigma^{ret}$.
This method is energy conserving.
We note that the propagation of $\omega$ and $\rho$ mesons at the
HADES energy range have been investigated in \cite{KampferWolf-2010,Almasi}
with the same method, where we have shown that the ``off-shell''
transport is consistent in the sense that mesons reaching the vacuum regain
their vacuum properties. Similar results were obtained for charmonium mesons in
\cite{WolfCharm_PhysLett}.

\section{Results}

In ref. \cite{WolfCharm_PhysLett} we have studied the dynamics of charmonium
production in antiproton induced reactions not very far from the threshold,
and the following plausible picture has been found:
Most of the antiprotons annihilate on, or close to the surface of the heavy
nucleus, creating a charmonium with a certain probability.
The charmonium travels
through the interior of the nucleus giving some contribution to the
dilepton yield. That is, the dileptons are treated perturbatively.
Traversing the thin surface again on the other side of the nucleus, it
arrives to the vacuum, where most charmoniums actually decay. We found
a two-peak structure, where one of the peaks comes from decays in the vacuum,
the other one from the decays inside the nucleus.
The distance of the two peaks corresponds to a mass shift at approximately
$0.9 \rho_0$ density. The D\,-\,meson loop contributes only $25-30$~MeV to
the mass shift. The rest (which is expected to be the major part) is the
result of the second order Stark effect, thus we can determine the gluon
condensate that has resulted in such a mass shift. The considered energy
regime will be available by the forthcoming PANDA experiment at FAIR. However,
there might be some difficulty to get quantitative conclusions since the
vacuum peak (originating from charmonium states created in the medium,
but decaying outside with their vacuum mass and width) is much
stronger than the one from the in-medium decays, so a good mass resolution of
the detector is required.
There may be another possibility to observe the mass shift, namely to measure
the excitation function of the $\Psi$(3686) state in antiproton induced
reactions. 

\begin{figure}[htb]
\centerline{%
  \includegraphics[width=0.7\textwidth]{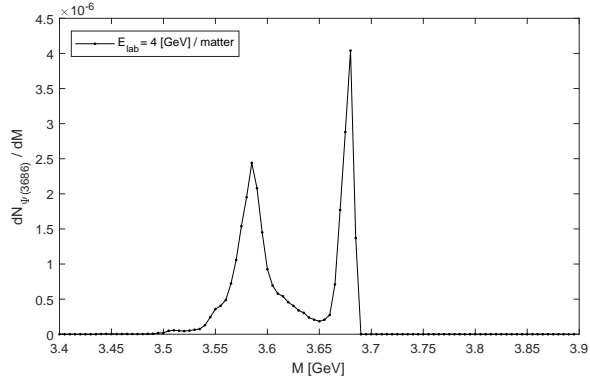}
}
\caption{$\Psi$(3686) contribution to the dilepton spectra taking into
  account the in-medium modifications.}
\label{Fig:Dilepspectra}
\end{figure}
In Fig.~\ref{Fig:Dilepspectra} we show the charmonium contributions to
the dilepton spectrum in a $\bar p$ Au 4 GeV beam kinetic energy collision.
We can observe the
same two peak structure as in \cite{WolfCharm_PhysLett}. In contrast to higher
energies, here the medium contribution is much larger compared to the vacuum
one. The reason is that very close to threshold the charmonium is much slower
than at higher energies, therefore, spend much more time inside the nucleus
increasing its contribution to the dilepton yield from the medium.

\begin{figure}[htb]
\centerline{%
  \includegraphics[width=0.7\textwidth]{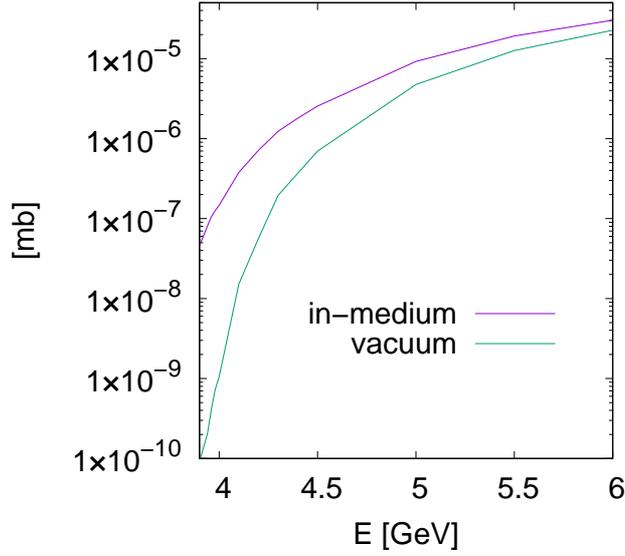}
}
\caption{Excitation function of the $\Psi$(3686) with in-medium and with
  vacuum properties.}
\label{Fig:Excit}
\end{figure}
In Fig \ref{Fig:Excit} we show the excitation function of the $\Psi$(3686)
production in $\bar p$ Au reactions using the propagation with
in-medium and with the vacuum properties. If we apply the in-medium mass-shift
and the collisional broadening of the $\Psi$(3686) its excitation function is
shifted downwards in energy drastically compared to the one calculated with
propagation
with its vacuum properties. Therefore, the mass shift can also be observed
by comparing the excitation function of the $\Psi$(3686) in $\bar p$ Au and in
$\bar p$ C, where the latter can be considered as the reference ``vacuum''
excitation function.

\section{Summary}

We calculated the charmonium contribution to the dilepton spectra. We
have shown that via their dileptonic decay there is a good chance to
observe the in-medium modification of the higher charmonium state:
$\Psi$(3686) in a $\bar p$ Au and $\bar p$ C collisions around the threshold
by measuring its excitation functions. This energy regime will be available by
the PANDA experiment.

\section*{Acknowledgments}

Gy.~W., M.~Z., G.~B., and P.~K. were supported by the Hungarian OTKA
fund K109462 and Gy.~W., M.~Z. and P.~K. by the HIC for FAIR Guest
Funds of the Goethe University Frankfurt, P. K. and M. Z. also acknowledge
support from the EMMI at the GSI, Darmstadt, Germany".

\end{document}